\documentclass[twocolumn,prd,showpacs,amsmath,amssymb]{revtex4}
\usepackage{graphicx,color}

\begin{document}
\title{Do stellar interferometry observations indicate UV modification of gravity? }

\author{Michael~Maziashvili}

\email{maziashvili@hepi.edu.ge}

\affiliation{
Institute of High Energy Physics, Tbilisi State University, 9 University
Str., Tbilisi 0186, Georgia }

\begin{abstract}

Recently it was demonstrated that the theoretical predictions for
the Planck scale space-time fluctuations derived by the
simultaneous consideration of quantum mechanics and general
relativity fail by many orders of magnitude against the stellar
interferometry observations. We propose the explanation of this
puzzle based on the short distance modification of gravity.
Roughly speaking the criterion of this modification to fit the
astronomical observations is that the gravitational radius of the
mass as large as $\sim 10^{26}$kg should not exceed $\sim
10^{-36}$cm. From this point of view we consider the modification
of Newtonian inverse square law due to Yukawa type correction and
the modification due to quantum gravitational corrections obtained
in the framework of nonperturbative RG approach. The latter one
does not satisfy the required criterion for modification.

\end{abstract}



\maketitle

\section{Introduction}

After the invention of quantum mechanics, the need to make it
consistent with special relativity led to creation of relativistic
quantum mechanics and quantum field theory subsequently \cite{LP}.
In (relativistic) quantum field theory, due to the possibility of
pair creation, a single particle cannot be localized more closely
than its Compton wavelength without losing its identity as a
single particle \cite{LP}. The vexing problem of unifying the
general relativity with quantum mechanics remains a holy grail of
theoretical physics. From the very outset one sees that there is
an important difference between the quantum mechanical space-time
and the general relativistic one. In quantum mechanics the
space-time is regarded as an essentially absolute, not affected by
the processes taking place inside it. While in general relativity
the space-time is given in terms of gravitational field that
undergoes some evolution during the physical process and therefore
becomes dynamical. Frequently we speak of length and time
intervals, i.e., space and time, without asking critically how one
actually measures them. It is clear that the coordinate system is
defined only by explicitly carrying out the space-time distance
measurements. Wigner and subsequently Salecker and Wigner
considered first this problem by applying simultaneously the
principles of quantum mechanics and general relativity \cite{WS}.
The analysis based on the \emph{gedankenexperiment} for space-time
measurement shows that the quantum fluctuations of geometry
becomes of the same order as the geometry itself at the Planck
scale resulting therefore in the notion of minimal observable
length beneath of which the geometrical properties of space-time
may be lost \cite{Garay}. This would give space-time a foam-like
structure at sub-Planckian distances, for instance it may show up
a fractal structure at this length scale while on larger scales it
would look smooth and with a well-defined metric structure. Planck
length, $l_p$, beneath of which the very concepts of space and
time lose their meaning may play a role analogous to the speed of
light in special relativity. If so, it means that Planck length
sets an observer independent minimum value of wavelength. A
quantum uncertainty in the position of a particle implies an
uncertainty in its momentum and due to gravity-energy interaction
leads to uncertainty in the geometry introducing in its turn an
additional position uncertainty for the particle. Due to
gravitational effects the minimum uncertainty in the position of a
particle with energy $E$ takes the form \[\delta x \gtrsim\mbox{
max}\left(E^{-1},~l_p^2E\right)~,\] showing therefore that the
Planck scale determines the lower bound for particle localization
no matter what its Compton wavelength, $E^{-1}$, is \cite{Garay}.
(Throughout this paper we set $\hbar=c=1$). By taking into account
the gravitational effects one can also argue that it is impossible
to accelerate a particle to post-Planckian energies \cite{CN}.

The above conclusions concerning the Planck
scale effects rely on the simultaneous consideration of usual principles of
quantum theory and general relativity. Loosely speaking in the framework of general relativity the Newtonian inverse square law is understood to be valid
for length scales from infinity to roughly the Planck scale. Thus far the direct experimental
tests for the short distance behavior of gravity are at sub-millimeter scale
\cite{HKH}, while , loosely speaking, quantum theory (electroweak interactions) is probed at distances
approaching $\sim 10^{-16}$cm. Therefore it is important to notice that the fundamental role
attributed to the Planck length is based on the assumption that the gravity is
unmodified over the $30$ orders of magnitude between where it is measured at
$\sim 10^{-2}$cm down to the Planck length $\sim 10^{-33}$cm. For the direct
test of gravitational interaction at short distances is highly nontrivial
problem, it is interesting therefore if indirect experimental measurements can
shed some light on the question and guide us to get some understanding of the gravity
behavior at relatively short distances. In what follows we discuss the UV
modification of gravity in light of the discrepancy between the stellar
interferometry observations and the theoretical predictions of Planck scale
space-time fluctuations demonstrated in \cite{LH, RTG}. We think UV
modification of gravity is most natural (and perhaps most exciting) possible
outcome of the puzzle.

\section{Space-time measurement}

Let us consider a \emph{gedankenexperiment} for distance
measurement proposed in \cite{Ma}. Our measuring device is
composed of a spherical clock localized in the region $\delta
x=2r_c$ and having a mass $m_c$, which also serves as a light
emitter and receiver, and a spherical mirror of radius $l/2$ and
mass $m$ surrounding the clock, ($r_c$ denotes the radius of the
clock), Fig.1. We are measuring a distance $l$ by sending a light
signal to the mirror under assumption that at the moment of light
emission the centers of clock and mirror coincide. However,
quantum uncertainties in the positions of the clock and mirror
introduce an inaccuracy $\delta l$ during the measurement.
(Throughout this paper we set $\hbar=c=1$).

Assuming the minimal uncertainty the spread in velocity of the clock
can be found as $ \delta v_c = \delta p_c/m_c =1/2m_c\delta x$ and analogously for the mirror as $ \delta v_m  =1/2(m+m_c)l$. After the
time $t = l$ elapsed by light to travel along the closed path
clock-mirror-clock the total uncertainty during the measurement
takes the form \[
\delta l = \delta x +{ t\over 2m_c\delta x} +{ t\over 2(m+m_c)l}~, \] which after
minimization with respect to $\delta x$ gives
\begin{equation}\label{minim}\delta x=\sqrt{{l \over 2 m_c}}~,~~~\Rightarrow~~~\delta l_{min}=\sqrt{{2l
\over  m_c}}+{1\over 2(m+m_c)}~.\end{equation} This uncertainty diminishes
with increasing masses of the mirror and clock. But the masses are limited by
the requirement the clock and the whole device as well not to collapse into the black hole. In other words
the size of the clock(mirror) should be greater than twice its gravitational
radius. So by equating the size of the clock, $\delta x$, to its gravitational
radius taken with factor $2$, one gets the maximum allowed mass of the clock
\[m_c^{max}=\left({l\over 32 l_p^4}\right)^{1/3}~,\] and then equating the
size of the measuring device, $l$, to its gravitational radius taken again
with factor $2$ one finds the maximal allowed mass of the mirror
\[m^{max}+m_c^{max}={l\over 4l_p^2}~.\]  In this way from Eq.(\ref{minim}) one gets the minimum
error in measuring a length $l$ to be
\begin{equation}\label{uncleng}\delta l_{min}=\left(16\, l_p^2\,l\right)^{1/ 3}+2l_p^{2}l^{-1}~.
\end{equation} From this equation it follows immediately that there exists the
minimal observable length of the order of $\sim l_p$. From Eq.(\ref{uncleng}) one readily gets the minimal uncertainty
in time measurement, $\delta t_{min}$, by replacing $l_p$ and $l$ with $t_p$
and $t$ respectively. Since the first term in Eq.(\ref{uncleng}) is much
greater than the second one for $l\gg l_p$ and they become comparable at $l\sim l_p$, one can say with no loss of generality
that for $l\gtrsim l_p$ our precision of space-time measurement is
limited by the measurement process itself such that
\begin{equation}\label{minlengunc}\delta l_{min}\gtrsim  2^{4/
3}\left (l_p^{2}l \right)^{1/ 3}~,~~~~\delta t_{min}\gtrsim 2^{4/
3}\left (t_p^{2}t \right)^{1/ 3}~.\end{equation}
Loosely speaking, in Eq.(\ref{uncleng}) one can consider the first and second
terms as the uncertainties contributed to the measurement by the clock and
mirror respectively. Due to maximal symmetry of the measuring device
considered here one can hope it provides minimal uncertainty in length
measurement.

\begin{figure}[t]
\includegraphics{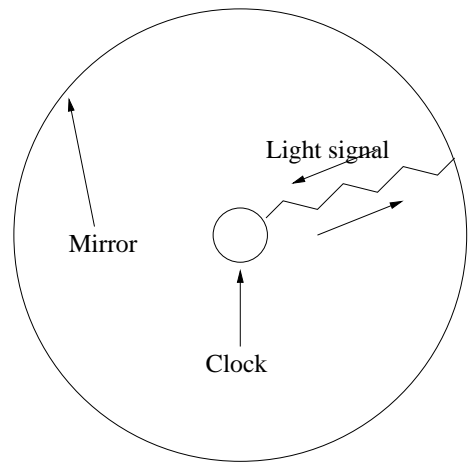}\\
\caption{ The device for distance measurement.}
\end{figure}

The above discussion to be self-consistent let us notice that due to
Eq.(\ref{minlengunc}) the wavelength of photon used for a measurement may be
known with the accuracy $\delta\lambda\sim (l_p^{2}\lambda)^{1/3}$ that
results in additional uncertainty \begin{equation}\label{totalucerlam}\delta
l_{min}^{(\lambda)}\gtrsim 2^{4/3}\left
(l_p^{2}l\right)^{{1/3}}+{l\over\lambda}\delta\lambda  \gtrsim 2^{4/3}\left [
\left (l_p^{2}l\right)^{{1/3}}+{l_p^{{2/ 3}}l\over \lambda^{{2/3}}}\right ]~.
\end{equation}(For the sake of simplicity we assume that the wave length of
the photon is not affected by the gravitational field of the
clock). For $\lambda$ can not be greater than $l$ in order to
measure this distance the latter term in Eq.(\ref{totalucerlam})
can be minimized by taking $\lambda\sim l$. In this case the
latter term becomes comparable to the first one and therefore
justifies the Eq.(\ref{minlengunc}). So, in general one can say
that our precision in space-time distance measurement is
inherently limited by the measurement process itself such that
\begin{equation*}\delta l_{min}^{(\lambda)}\gtrsim 2^{4/3}{l_p^{{2/
3}}l\over \lambda^{{2/3}}}~,~~~~~~\delta
g_{\mu\nu}^{(\lambda)}\gtrsim
2^{4/3}\left({l_p\over\lambda}\right)^{{2/ 3}}~. \label{fluc}
\end{equation*} The quantum with momentum $p$ has the wavelength
$\lambda= 2\pi p^{-1}$ and from Eq.(\ref{minlengunc})   For the
energy-momentum uncertainties one gets

\begin{equation}\label{enmomun} \delta
p\gtrsim { 2^{4/3} p^{5/3}\over (2\pi)^{2/3}m_p^{2/3}} ~,~~~~~~\delta E=pE^{-1}\delta p~.  \end{equation}

\section{Astronomical observations }

An interesting idea proposed
in \cite{LH} is to consider the phase incoherence of light coming to
us from extragalactic sources. Since the phase
coherence of light from an astronomical source incident upon a two-element
interferometer is necessary condition to subsequently form interference
fringes, such observations offer by far the most sensitive and uncontroversial
test. The interference pattern when the source is viewed through a
telescope will be destroyed if the phase incoherence, $\delta \varphi$, approaches $2\pi$. In other
words, if the light with wavelength $\lambda$ received from a celestial
optical source located at a distance $l$ away produces the normal interference
pattern, the corresponding phase uncertainty should satisfy the condition
\begin{equation*} \delta\varphi < 2\pi~.  \end{equation*}

The energy-momentum uncertainties, Eq.(\ref{enmomun}), result
in uncertainties of phase and group velocities of the photon \[\delta v_p=2{\delta E\over E}~,~~~~~~\delta v_g={d\delta
E\over dp}~.\]
The light with wavelength $\lambda$ traveling
over a distance $l$ accumulates the phase uncertainty
 \begin{equation} \label{phaseincoh} \delta\varphi= {2\pi l\over\lambda}\delta\left({v_p\over v_g}\right)  = {2\pi l\over\lambda}(\delta v_p+\delta
v_g)~,  \end{equation} where $\delta(v_p/v_g)$ stands for maximal
deviation of the ratio $v_p/v_g$ due to fluctuations $\delta
v_p,~\delta v_g$ taking a $\pm$ sign with equal probability. Using
the Eqs.(\ref{enmomun},\ref{phaseincoh}) one finds
\begin{equation} \label{intercond}   \delta \varphi \approx
58.0513\, {l_p^{2/3}l\over \lambda^{5/3}} ~.\end{equation}

Consider the examples used in \cite{LH}. The Young's type of interference effects were clearly seen
at $\lambda = 2.2 \mu$m light from
a source at $1.012$ kpc distance, viz. the star S Ser,
using the
Infra-red Optical Telescope Array, which enabled a radius determination of
the star \cite{BTG}. Using these values of wavelength and distance from Eq.(\ref{intercond}) one gets \[\delta\varphi\approx
6.74\times 10^{20/3}~. \]

Airy rings  (circular diffraction) were clearly visible
at both the zeroth and first maxima in an observation
of the active galaxy PKS1413+135 ($l = 1.216$ Gpc) by
the Hubble Space Telescope at $ 1.6 \mu $m wavelength \cite{PSC}.
Correspondingly, using the Eq.(\ref{intercond}) one gets the phase incoherence
\[\delta\varphi\approx 1.37\times 10^{41/3}~. \]

This idea was further elaborated in \cite{RTG} by observing that the uncertainty in length measurement will lead to
apparent blurring of distant point sources observed through the telescope
confirming the above huge discrepancy between the theoretical predictions and the
astronomical observations.

\section{Discussion}

From the preceding section one sees that the theoretical
prediction for minimum lengths uncertainty in measuring the length
of the order of $\sim \mu$m is about $14$ orders of magnitude
greater in comparison with astronomical observations. The extent
to which a rigorous physical meaning can be attributed to
Eq.(\ref{minlengunc}) must be founded on a direct appeal to
experiments and measurements. The only ingredients used in
derivation of minimum length uncertainty are position-momentum
uncertainty relation and the standard expression of gravitational
radius. Therefore the contradiction between theoretical results
and experiments can be understood as a failure of at least one of
these ingredients. From section II one sees that the size and the
mass of the clock giving the minimal uncertainty in measuring a
distance $l$ are given by \[r_c \sim l_p^{2/3}l^{1/3}~,~~~~~~m_c
\sim l_p^{-4/3}l^{1/3}~.\] By taking into account that $r_c$ is
comparable to the gravitational radius of the clock evaluated on
the bases of inverse square law and experimentally established
lower limit to date for the Newtonian inverse square law is $\sim
10^{-2}$cm, one can say for sure that the Eq.(\ref{minlengunc})
describes physical reality for $l \gtrsim 10^{60}$cm but can not
be considered as rigorously established result beneath this length
scale. In analyzing astronomical observations considered in \cite{
LH, RTG} we are evaluating the uncertainty for the photon
wavelength being of the order of $\sim \mu$m by means of
Eq.(\ref{minlengunc}) assuming thereby the validity of Newtonian
inverse square law and the Heisenberg position-momentum
uncertainty relation at the lengths scale $\sim 10^{-23}$cm.
Simply speaking the position-momentum uncertainty relation is
based on the de Broglie relation plus simple properties common to
all waves. If we want to measure the localization of some particle
with accuracy $\delta x$ we need to use the quantum (photon for
instance) the wavelength of which $\lambda \lesssim \delta x$. Due
to de Broglie relation, such a quantum is characterized with
momentum \[p \sim \lambda^{-1},\] and during the scattering on the
particle being observed imparts to it the momentum $\delta p$
evaluated roughly as $\delta p\sim p $ leading thereby to the
position-momentum uncertainty relation
\[\delta x \,\delta p \gtrsim 1~.\] The length scale $\sim
10^{-23}$ is corresponding to the ultra-high energy cosmic rays,
let us assume the position-momentum uncertainty relation is valid
at least up to this scale. If position-momentum uncertainty
relation is maintained, then as a weak spot in evaluation of
minimum length uncertainty should be considered the use of
Newtonian inverse square law at the length scale $\sim
10^{-23}$cm.

Considering the modified gravity law
\[g_{00}=g_{rr}^{-1}=1-{f(r,~m)\over r}~,\] for the gravitational
radius one finds
\begin{equation}\label{modgravrad} r_g=f(r_g,~m)~.\end{equation} As it was discussed in section II the maximal allowed mass of the clock
 is determined through the equation

\begin{equation}\label{maxclmasmod}\sqrt{{l\over 2m_c^{max}}}=2r_g~,\end{equation} which together with the Eq.(\ref{modgravrad})
determines gravitational radius of the clock giving the minimal
error in measurement of a given length $l$ as a function of $l$
 \begin{equation}\label{optclgravrad}r_g=f(r_g,~l/8r_g)~.\end{equation}
Knowing the maximal allowed mass of the clock one can determine the maximal
allowed mass of the mirror as a function of $l$ through the equation
\begin{equation*} l=2f\left(l/2,~m^{max}+m_c^{max}\right )~,  \end{equation*}
and estimate the minimum length uncertainty by means of Eq.(\ref{minim}) as

\begin{equation}\label{lenguncmodgrav}\delta l_{min} =4r_g(l)+{1\over 2\left(m^{max}+m_c^{max}\right)}~, \end{equation}
where $r_g(l)$ is the solution of Eq.(\ref{optclgravrad}). Let us omit the
second term in Eq.(\ref{lenguncmodgrav}) which decreases with $l$ and for
the length scale $\sim\mu$m, which is the case in the above considered experiments, can be
neglected in comparison with the first term. Then for the
phase and group velocity uncertainties of the photon one gets \[\delta
v_p={8r_g(\lambda)\over \lambda}~,~~~~\delta
v_g={8r_g(\lambda)\over\lambda}-4{dr_g(\lambda)\over d\lambda}~.\]
Correspondingly the phase incoherence accumulated along a distance $l$ takes
the form \begin{equation}\label{phaseincmodg}\delta\varphi={2\pi l\over\lambda}\left\{ {8r_g(\lambda)\over
\lambda}+\left| {8r_g(\lambda)\over\lambda}-4{dr_g(\lambda)\over d\lambda}
\right| \right\}~. \end{equation}

This equation combined with the second example of astronomical observation
(HST image of PKS1413+135) puts the following limit \[r_g(\lambda =
1.6\mu\mbox{m}) < 0.85\times 10^{-36}\mbox{cm}. \] From Eq.(\ref{maxclmasmod})
one sees that the clock mass corresponding to this gravitational radius satisfy
\[m_c > 0.12 \times 10^{27}\mbox{kg}~. \] Besides this restriction on
$r_g(\lambda)$, from Eq.(\ref{phaseincmodg}) one sees that
$dr_g(\lambda)/d\lambda$ is also subject to the constraint due to experimental
observations.

To illustrate the discussion let
us consider the running Newtonian constant of the form
\begin{equation}\label{potential}G(r)=l_p^2\left(1-\alpha
e^{-r/r_0}\right)~,\end{equation} the direct experimental test for
which in the case $|\alpha|\geq 1$ sets $r_0 < 197\times
10^{-4}$cm \cite{HKH}. Correspondingly, the equation for the
gravitational radius takes the form
\begin{equation*} r_g=2l_p^2m\left(1-\alpha e^{-r_g/r_0}\right)~. \end{equation*} We notice that
the Yukawa type correction to the Newtonian inverse square law
with negative $\alpha$ is inconsistent for our purposes. From the
above discussion one finds that gravitational radius of the clock
giving the minimal error in measuring of a given length $l$ is
determined by the equation
\begin{equation*}  r_g^3-{l_p^2l\over 4}\left(1-\alpha e^{-r_g/r_0}\right)=0~.  \end{equation*}
Taking the upper bounds for the parameters
$r_g(\lambda=1.6\mu\mbox{m}),~r_0$, for parameter $\alpha$ one
finds \[\alpha = \left(1-{4r_g^3\over
l_p^2\lambda}\right)e^{r_g/r_0} \approx \left(1-0.58\times
10^{-38}\right)e^{0.43\times 10^{-34}}~.\] So, in this case
$\alpha$ equals $1$ with a great accuracy. Denoting $r_g'\equiv
dr_g/dl$ one finds
\begin{equation}\label{derofgravrad}r_g'={4r_0r_g^3\over l\left(12\,r_0r_g^2-\alpha
l_p^2le^{-r_g/r_0}\right)}~. \end{equation} Using the above
parameters, from Eq.(\ref{derofgravrad}) one gets
\[r_g'(\lambda=1.6\mu\mbox{m}) \approx 72.06 \times 10^{-38}~.\]
Therefore the running Newtonian constant (\ref{potential}) for
$\alpha \approx 1$ satisfies the experimental constraints on
Eq.(\ref{phaseincmodg}).

Another example of running Newton constant obtained by means of
the Wilson-type effective action has the form \cite{BR}

\begin{equation*}\label{runnewcon}G(r)={l_p^2r^3\over
r^3+2.504l_p^2\left(r+4.5l_p^2m\right)}~,\end{equation*} where $m$
is the mass of the source. By taking the quantum corrected
equation for gravitational radius as \cite{BR}
\[r_g=2G(r_g)m~,\] one finds the following expression for outer
horizon
\begin{widetext}
\begin{eqnarray}\label{ograd}r_g/l_p^2m&=&0.667 +
 0.265  \left(16 - 139.5\,t +10.392\sqrt{t(t-0.204 )(176.392 +
 t)}\right)^{1/3} \nonumber\\&-& {0.41(3t-4)\over \left(16 - 139.5 \,t +
10.392\sqrt{t(t-0.204)(176.392+t)}\right)^{1/3}}~,\end{eqnarray}\end{widetext}
where $t\equiv 2.504m_p^2/m^2$. (An interesting feature of this
running Newton constant is that there is a critical value of mass,
$t_{cr}=0.204,~~~ m_{cr}=3.503\,m_p$, below which the horizon
disappears and leads therefore to the black hole remnants of about
Planck mass \cite{BR, BRM}.) From Eq.(\ref{ograd}) one sees that
this running Newtonian constant gives practically the standard
value of gravitational radius for $m \sim 10^{26}$kg and therefore
does not satisfy the criterion for explaining the astronomical
observations of the preceding section.

The basic conceptual point coming from the
\emph{gedankenexperiment} for space-time measurement is that
quantum fluctuations of measuring device in the process of making
measurement does not allow the space-time intervals to be
precisely determined. In other words the space-time is disturbed
by the measurement process itself limiting thereby our precision
in space-time measurement. Therefore, one should take note of
difficulty in evaluating of space-time disturbance during the real
measurement stemming from the fact that in the real measurements
we are not using optimal devices minimizing the space-time
uncertainties caused by the measurement process itself.

\vspace{0.2cm}
\begin{acknowledgments}

The author is greatly indebted to David Langlois for cordial hospitality at
APC (Astroparticule et Cosmologie, CNRS, Universit\'e Paris 7) and IAP
(Institut d'Astrophysique de Paris), where this work was done. Thanks are due
to L.~Sorbo and A.~Fabbri for useful conversations. The work was supported by
the \emph{INTAS Fellowship for Young Scientists}, the
\emph{Georgian President Fellowship for Young Scientists} and the
grant \emph{FEL. REG. $980767$}.
\end{acknowledgments}


\begin{thebibliography}{99}


\bibitem{LP}
L.~Landau and R.~Peierls,, Z. Phys. 69 (1931) 56, Reprinted with the English
translation in \emph{Collected Papres of L.~D.~Landau} (ed. D. ter Haar,
Pergamon Press, Oxford, 1965); V.~Berestetskii, E.~Lifshitz and L.~Pitaevskii,
\emph{Quantum Electrodynamics}, (Landau and Lifshitz Course of Theoretical
Physics, Pergamon Press, Oxford, 1982).


\bibitem{WS}
E.~Wigner, Rev. Mod. Phys. 29
(1957) 255; H.~Salecker and E.~
Wigner, Phys.
Rev. 109 (1958) 571; F.~Karolyhazy, Nuovo Cim. A42 (1966) 390; Y.~J.~Ng and H.~van~Dam, Mod. Phys. Lett. A9
(1994) 335.


\bibitem{Garay}
 L.~Garay, Int. J. Mod. Phys. A10 (1995)
145, gr-qc/9403008.

\bibitem{CN}
A.~Casher and S.~Nussinov, hep-th/9709127.

\bibitem{HKH}
 C.~Hoyle, D.~Kapner, B.~Heckel, E.~Adelberger, J.~Gundlach, U.~Schmidt and
H.~Swanson Phys. Rev. D70 (2004) 042004, hep-ph/0405262.

\bibitem{LH}
R.~Lieu and L.~Hillman, astro-ph/0211402; R.~Lieu and L.~Hillman, Astrophys.
J.585 (2003) L77, astro-ph/0301184.

\bibitem{RTG}
R.~Ragazzoni, M.~Turatto and W.~Gaessler, Astrophys. J. 587 (2003) L1, astro-ph/0303043.

\bibitem{Ma}
M.~Maziashvili, hep-ph/0605146.

\bibitem{BTG}
G.~van Belle, R.~Thompson and M.~Creech-Eakman, Astron. J. 124 (2002) 1706, astro-ph/0210167.

\bibitem{PSC}
E.~Perlman et al., Astron. J. 124 (2002) 2401, astro-ph/0208167.

\bibitem{BR}
A.~Bonanno and M.~Reuter, Phys. Rev. D62 (2000) 043008,
hep-th/0002196.

\bibitem{BRM}
A.~Bonanno and M.~Reuter, Phys. Rev. D73 (2006) 083005,
hep-th/0602159; M.~Maziashvili, Phys. Lett. B635 (2006) 232,
gr-qc/0511054.




\end{thebibliography}
\end{document}